\documentclass[%
 aip,
 amsmath,amssymb,
 reprint,%
]{revtex4-1}

\usepackage{graphicx}
\usepackage{dcolumn}
\usepackage{bm}

\usepackage{mathptmx}
\usepackage{color}
\usepackage{todonotes}
\usepackage{upgreek}

\begin{document}

\preprint{AIP/123-QED}

\title{Vectorial light-matter interaction \\ -- exploring spatially structured complex light fields}

\author{Jinwen Wang}
\affiliation{School of Physics and Astronomy, University of Glasgow, G12 8QQ, United Kingdom}
\affiliation{Shaanxi Key Laboratory of Quantum Information and Quantum Optoelectronic Devices, School of Science, Xi'an Jiaotong University, Xi'an 710049, China}
\author{Francesco Castellucci}%
\author{Sonja Franke-Arnold}
 \email{Sonja.Franke-Arnold@glasgow.ac.uk}
\affiliation{School of Physics and Astronomy, University of Glasgow, G12 8QQ, United Kingdom}

\date{\today}

\begin{abstract}
Research on spatially-structured light has seen an explosion in activity over the past decades, powered by technological advances for generating such light, and driven by questions of fundamental science as well as engineering applications.  In this review we highlight work on the interaction of vector light fields with atoms, and matter in general. This vibrant research area explores the full potential of light, with clear benefits for classical as well as quantum applications.
\end{abstract}
\maketitle

\section{Introduction} \label{sectionIntro}
Considering the vector nature of light is relevant in all physical systems that are affected by interference in its many guises, including diffraction, scattering and nonlinear processes, and by interaction with matter that has a polarization-sensitive symmetry.  Being able to design the polarization structure of optical beams and even of individual photons opens new opportunities in the classical as well as the quantum regime. This includes the study of topological phenomena, \cite{lu2014topological,ozawa2019topological} the conversion between phase and polarization singularities, \cite{dennis2009singular,marrucci2011spin} and interaction between orbital and spin angular momentum, \cite{bliokh2015spin} as well as technological advances in polarimetry and ellipsometry, sensing and focussing beyond the conventional diffraction limit. \cite{youngworth2000focusing,dorn2003sharper}

Classical vector fields can mimic quantum behavior in their correlation between polarization and spatial degree of freedom, and carry an increased information content compared to homogeneously polarized light. \cite{holleczek2011classical} Both of these features make them interesting candidates for multiplexing in communication and information systems. \cite{erhard2018twisted,cozzolino2019high} A wide range of research and review articles provide information on the generation, properties, as well as classical and quantum applications of vector beams. \cite{rosales2018review,chen2018vectorial,forbes2019quantum}

In this review article we examine current work on the interaction of vector light with matter, and specifically with atomic gasses. This is a young research area, and the majority of experimental research falls within the semi-classical regime. The interaction of structured light with matter is, however, relevant for all elements of quantum information networks -- for the transfer, storage and manipulation of high dimensional quantum information. This includes passive processes, such as the propagation of light through turbulence or density fluctuations, the effect of dichroism and birefringence, and the desired or unwanted mode conversion due to linear and nonlinear interaction. 

Atoms, on the other hand, are active optical elements: while the complex vector structure of a light beam is modified by its interaction with an atomic medium, the atomic populations and coherences are correspondingly modified by the optical beam -- effectively entangling optical with atomic structures.  Atomic interactions can be drastically enhanced in the vicinity of atomic resonances, leading to significant nonlinear effects, with a response that can furthermore be altered by external magnetic fields.

Atomic transitions are intrinsically sensitive to the vector nature of light. The dominant effect in light-atom interaction, the electric dipole interaction, explicitly depends on the alignment of the optical field with the atomic dipole, affecting selection rules and transition strengths. 
Light polarization therefore plays an important role in the preparation, manipulation and detection of atomic states. Light polarization affects incoherent processes via optical pumping, as well as coherent parametric processes involving several atomic transitions. 

The first decades of exploring light-matter interaction have almost entirely been restricted to the study of  homogeneously polarized light, or `scalar' light  -- and indeed polarization structures tend to play little role in linear paraxial optics.    
The most prevalent, but maybe also least recognised use of spatially varying polarization is Sisyphus (or polarization gradient) cooling, honoured in the 1997 Nobel Prize. In this case the optical polarization is modulated along the beam propagation direction generated by a pair of  counter-propagating laser beams.  This generates a modulated AC Stark shift which, in conjunction with optical pumping, allows the dissipation of the dissipation of energy from the atomic motion to the optical field.  While this is a well-understood process, of benefit to any experimenter wishing to obtain lower atomic temperatures from a magneto-optical trap, it is likely still hiding some secrets. 

In this review article, however, we will concentrate on transverse polarization structures in light, and describe their interaction with atomic as well as other nonlinear media.  We start with a brief description of vector light fields in Section \ref{sectionFields}, discuss in Section \ref{sectionDichroism} how they exert optical dichroism and birefringence  in matter, and in  the propagation of vector beams through nonlinear media. The next two sections concentrate on situations where vector light interacts coherently with atomic or other media, allowing parametric processes which facilitate mode conversion (Section \ref{sectionMode}) and modify the medium itself (Section \ref{sectionEIT}). We briefly discuss the effects of vector fields under strong focussing in Section \ref{sectionFocus}, before our Conclusion.

\section{Spatially structured 
vector fields}
\label{sectionFields}
\begin{figure*}[!t]
\centering
\includegraphics[width=17cm]{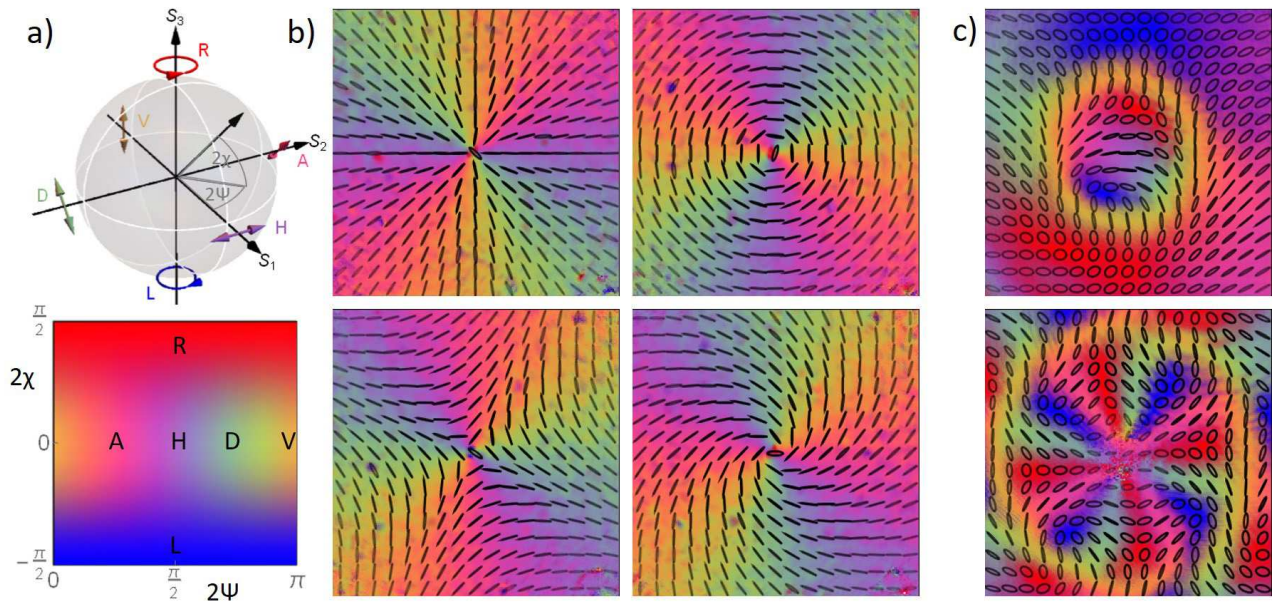}
\caption{Vector beams. a) Polarization Poincar{é} sphere and associated colormap, linking each Stokes vector to a unique color. b) Experimental images of radial, azimuthal and two hybrid polarization. These polarization structures together with homogeneous right and left circular polarization, can be used as an alternative mutually unbiased vector basis. c) two examples of experimental Poincar{é} beams. 
\label{fig_PoincarePol}}
\end{figure*}
Historically, structured light referred to a modulation of the intensity profile of a light beam, implemented by amplitude filters. This was later extended to include tailored complex amplitude profiles, which can be conveniently implemented with programmable diffractive elements including spatial light modulators (SLMs), \cite{forbes2016creation} digital micromirror devices (DMDs), \cite{selyem2019basis,manthalkar2020all,zhao2020determining} or various alternative methods. \cite{intaravanne2020recent} The most prominent example of such spatially tailored complex light are the orbital angular momentum (OAM) modes, e.g. Laguerre-Gauss  \cite{allen1992orbital,yao2011orbital} or Bessel modes, \cite{mcgloin2005bessel,fahrbach2010microscopy}  and recently also  Ince–Gaussian vector modes with elliptical symmetry. \cite{otte2018sculpting,yao2020classically}

Light, or at least a propagating paraxial light beam, is a transverse vector field with two independent polarization components. In order to fully tailor vector fields, one needs to  control the complex amplitudes of each polarization component separately.\cite{zhan2009cylindrical,rubinsztein2016roadmap}  Doing so generates spatially varying polarization states, including radial, azimuthal, spiral \cite{milione2011higher} and hybrid polarization, \cite{lerman2010generation,wang2010new} as well as full Poincar\'e beams \cite{beckley2010full,beckley2012full,cardano2013generation,ling2016characterization,arora2019full} and custom-designed polarization modes. \cite{turpin2015polarization,li2016generation,alpmann2017dynamic,li2018generation,liu2018highly,li2018polarization,chang2019tunable,li2019shaping,gu2019generation,wang2020complete}  

A paraxial vectorial light beam can be written as 
\begin{eqnarray}
\vec{u}(\vec{r}_\perp) & = &\begin{pmatrix} u_h(\vec{r}_\perp)\exp[i \Phi_h(\vec{r}_\perp)]  \\u_v(\vec{r}_\perp) \exp[i \Phi_v(\vec{r}_\perp)] \end{pmatrix} 
\end{eqnarray}
where $u_{h,v}(\vec{r}_\perp)$ and $\Phi_{h,v}(\vec{r}_\perp)$ are the spatial amplitudes and phases of the horizontal and vertical polarization components, and $\vec{r}_\perp=(x,y)$ is the transverse position. In the following we will omit the explicit position dependence for clarity. We can rewrite this equation to accentuate the physical interpretation as
\begin{eqnarray}
\vec{u}(\vec{r}_\perp)&=& u \exp[i \Phi] \begin{pmatrix} \cos(\Theta)\exp[i \,\Delta\Phi] ) \\\sin(\Theta) \end{pmatrix}  \end{eqnarray}
Here $u=\sqrt{u_h^2+u_v^2}$ is the position dependent complex amplitude of the light, and $\Phi=\Phi_v$ the overall spatially varying phase. The third term is the local polarization vector, with an ellipticity determined by the differential phase, $\Delta\Phi=\Phi_h - \Phi_v$, and an orientation against the horizontal axis given by $\Theta$ which can be found from the amplitude ratio: 
$\tan \Theta = u_v/u_h.$
We note that instead of a decomposition into a linear polarization basis we could have chosen a rotational basis, or in fact any other orthonormal basis set. The local polarization of such beams varies from point to point and is usually measured by determining the associated Stokes parameters.  

We have at our disposal a plethora of methods to generate structured vector fields. These fall into two broad categories: the generation of specific and of arbitrary modes. The former include the use of birefringent or dichroic materials within laser cavity as active elements, \cite{naidoo2016controlled,forbes2019structured,he2019complex} or outside as passive elements,  \cite{zhan2009cylindrical, liu2016} 
s- and q-plates, \cite{marrucci2006optical,rubano2019q} plasmonic metasurfaces \cite{Yue2016,intaravanne2020recent} and Fresnel cones. \cite{radwell2016achromatic,hawley2020tight}  These methods tend to be highly efficient, but the spatial modes are restricted to a specific 2D subset within the infinite-dimensional spatial state space, usually with the same amplitude but different phase profiles. A typical example would be the generation of modes $\propto\exp( i \varphi) \hat{\sigma}_{-} +\exp(- i \phi) \hat{\sigma}_{+} $  where $\varphi$ denotes the azimuthal angle, and $\hat{\sigma}_{\pm}$ are left- and right-handed polarization states. Polarization optics then allow manipulations within this subset, generating radial, azimuthal, hybrid and spiralling polarization states. The resulting modes are in general not eigenmodes of propagation and experiments have to be performed in the appropriate imaging plane.

Generating arbitrary structured vector beams, instead, requires the independent design of the complex amplitudes of the horizontal and vertical polarization (or any other independent polarization components), without disturbing the transverse coherence of the light field. This can be achieved by placing programmable devices like SLMs or DMDs within interferometers. \cite{Niziev2006,maurer2007tailoring,milione2012higher,Otte2016,selyem2019basis} These techniques allow on-demand and real-time structuring of arbitrary vector light fields, \cite{Mitchell2016} limited only by the spatial and temporal resolution of the beam-shaping device. The drawback of these methods is that they operate at low efficiencies, and hence are more suited for classical light beams rather than single photons.

Fig.~\ref{fig_PoincarePol} illustrates the wide range of polarization structures, showing a selection of vector beams generated in our lab using a DMD based generation method described in a previous publication.\cite{selyem2019basis} 
In a) we explain the colorscheme used throughout this review to depict polarization profiles. 
Fig.~\ref{fig_PoincarePol}b) shows a subsection of the mode family built from the Laguerre-Gauss modes LG$_p^{\ell}=$LG$_0^0$ and LG$_0^2$. The obtained polarization profiles, incidentally, are identical to the more familiar radial, azimuthal and hybrid polarization resulting from superpositions of LG$_0^{\pm 1}$, but with a difference in the overall phase of  $\exp(-i \varphi)$: the depicted beams carry a net OAM of $\hbar$ per photon, in contrast to the balanced case without OAM. Transformations within this mode family can easily be realised by mirror reflections and wave retarders. The polarization mode set of Fig.~\ref{fig_PoincarePol}b in combination with homogeneous right and left circular states form a rotationally symmetric set of unbiased basis, with applications in alignment-free quantum communication. \cite{d2012complete}  Fig.~\ref{fig_PoincarePol}c shows two examples of Poincar\'e beams, each containing the complete range of polarizations across their profile, effectively mapping the polarization Poincar\'e sphere onto the transverse beam profile. The particular examples shown here are the mode LG$_0^1 \hat{h}$+LG$_1^{-1}\hat{v}$ (top), and LG$_0^2 \hat{h}$+LG$_1^{-3}\hat{v}$ (bottom).

\section{Dichroism and birefringence} 
\label{sectionDichroism}

Many materials, crystals, fluids, and also atomic media, can affect the polarization of light, either by optical dichroism or birefrigence.  The former stems from polarization-dependent absorption, and the latter from polarization-dependent dispersion, and each may discriminate between linear or circular polarizations. Traditionally these effects are observed for homogeneously polarized light beams, but of course they become even more interesting for polarization structured light. As absorption and dispersion are linked via Kramers-Kronig relations, dichroism and birefringence are connected, which is especially noticable for near-resonant excitation in atomic media. 

Birefringence and dichroism can occur naturally in crystals or be induced by external forces. Crystals that naturally feature such an anisotropy are widely used to fabricate many kinds of polarization optics, including polarizers and wave retarders, and many publications describe the propagation of complex vector beams in anisotropic crystals. \cite{cincotti2003radially,flossmann2005polarization,li2012paraxial,li2013propagation,lian2017polarization,gu2018experimental,xie2018paraxial,guo2018dynamic,su2020evolution,khonina2020variable} 

Of particular importance for the manipulation of vector light fields are liquid crystals, which form the building blocks of SLMs and q-plates. The spatially varying birefringence of q-plates can be used as an interface to convert between spin and orbital states, for classical beams as well as for individual photons. \cite{nagali2009quantum}  
The flexibility of this approach has recently been beautifully demonstrated by generating tunable two-photon quantum interference of vector light, measured by observing the Hong-Ou-Mandel dip. \cite{d2019tunable}

Atoms are not naturally anisotropic, but they can become polarization sensitive in external fields, most notably by magneto-optical effects and by optical pumping in strong probe light. These effects are well understood and utilised in atomic magnetometry \cite{budker2002resonant,budker2007optical,kitching2011atomic,kitching2018chip} and polarization selective absorption spectroscopy. \cite{lee1994zeeman,harris2006polarization,moon2008effect,do2008polarization}

Experiments based on optical pumping usually use a strong pump laser to induce a spin alignment of the atomic medium, which is then tested with a co- or counter-propagating weak probe laser. The polarization of the strong pump determines the quantization axis and hence the spin alignment of the atoms, which modifies the interaction with the probe beam. \cite{nakayama1997optical}
Linear polarization along the quantization axis drives $\pi$ transitions, while linear polarization perpendicular to the quantization axis drives superpositions of ${\sigma}_{+}$ and ${\sigma}_{-}$ transitions, where their phase difference is determined by the orientation of the linear polarization within the perpendicular plane. \cite{auzinsh2010optically} 
\begin{figure*}
\centering
\includegraphics[width=16cm]{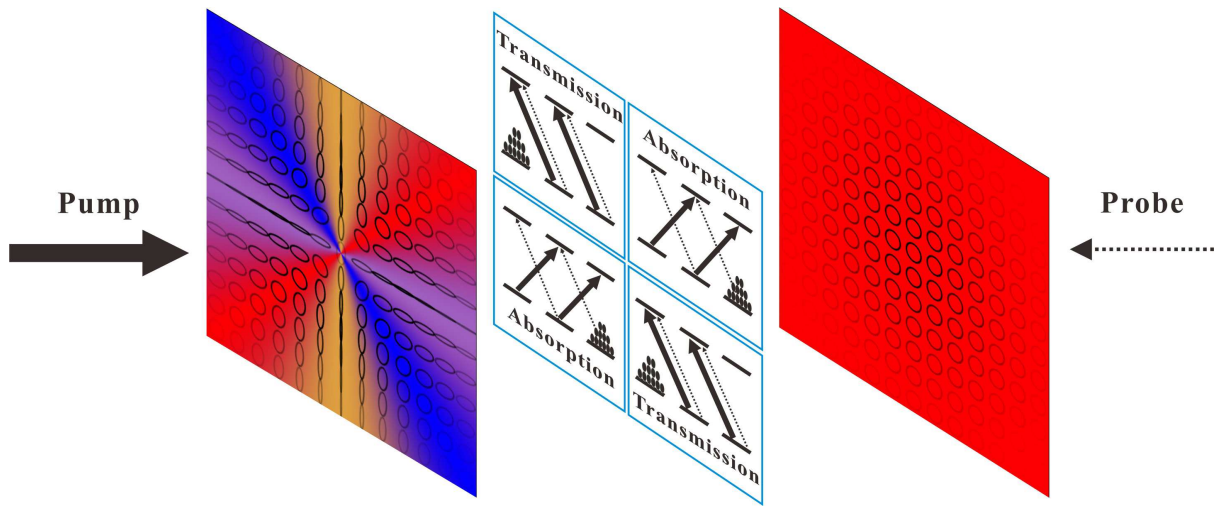}
\caption{Dichroism in an atomic medium induced by a vectorial pump field. If atoms are exposed to a strong pump beam with spatially varying polarization (left), optical pumping differs locally.  This induces spatially varying atomic population distributions (center) which can be probed by a weak uniform beam (right). In areas where the helicity of the probe matches that of the pump transmission is enhanced, whereas areas with opposite helicity are absorbed -- effectively realising spatially dependent absorption spectroscopy. 
}
\label{opticalpumping}
\end{figure*}

One of the first attempts at connecting the polarization structure of an optical beam to the spatial profiles of atomic spin alignments was published in a single-author paper in 2011.\cite{Fatemi2011} Experiments were performed in a moderately heated vapour cell of $^{85}$Rb, with a right hand circularly polarized Gaussian pump beam (i.e. $\hat{\sigma}_{-}$), and a much (two orders of magnitude) weaker 
co-axial counter-propagating vector vortex beam (VVB), 
driving the D2 line from F=2 $\to$ F'=2. Atoms were optically pumped into the magnetic sublevel $m_F=-2,$ thereby enhancing the transmission of the probe, where the local helicity of the probe matched that of the pump, and reduced where its helicity was opposite.  It is important to remember, of course, that optical polarization is denoted with respect to the beam's propagation direction, while atoms respond to the optical helicity, defined with respect to a set quantization direction. The optically-pumped atoms effectively display light-induced circular dichroism, and behave like a circular polarizer. 

Homogeneous pump beams result in homogeneous dichroism, whereas polarization-structured pump beams display a varying degree of dichroism, as illustrated in Fig.~\ref{opticalpumping}. If the role of pump and probe are reversed, a strong vector vortex beam will pump atoms into different magnetic sublevels depending on its local polarization direction, which then can be tested with a homogeoneously polarized probe, \cite{Fatemi2011} transferring spatially resolved information from pump to probe. 

Similar mechanisms can also be realised in more sophisticated level structures, with pump and probe addressing different atomic transitions via the Doppler effect. This was demonstrated e.g.\ on the D2 line of $^{87}$Rb using the crossover transition signal between $F=1 \to F'=1$ and $F=1 \to F'=0$. \cite{Li2019,wang2020optically} By combining a pump beam with (quasi) uniform amplitude and spatially varying polarization profile and a probe with spatially varying amplitude and homogeneous polarization, this configuration can be used for spatially-resolved optical information selection. \cite{Wang2018} 
The transverse profile of a probe beam with uniform circular polarization was encoded by an SLM with an image of spatially separated numbers. This probe was passed through an atomic vapor pumped by a strong hybrid polarized beam. The different information written in the probe beam could then be filtered by rotating the probe beam polarization with respect to the pump beam polarization distribution. Finally, mixed linear or circular dichroism can also be tested by using vector vortex beams for both pump and probe beams.\cite{Yang2019oe} The spatial  polarization profile of vector vortex beams leads to spatially varying  dichroism of the atoms and provides the opportunity for spatially tailored manipulation of light-atom interaction, which furthermore could be modulated by varying the detuning and intensity balance between pump and probe. 

It was shown that this pump-probe technique provides a direct tool for acquiring both SAM and OAM of structured light, \cite{Wang2019} analogous to projective measurements using a combination of waveplates and polarizers. The advantage of this technique is that the extracted part still maintains the original polarization and the vortex phase.
In other words, a circular polarizer or filter can be produced by manipulating the polarized state of atoms through a pump field. This scheme also provides the possibility of developing atom-optical devices and integrated devices based on atoms for projective measurements. 

An alternative method to induce anistropy within atomic media is via external magnetic fields. A field along the quantization axis shifts the Zeeman sublevels, changing the resonance frequencies for opposite circular polarized light components. 
The resulting difference in refractive indices leads to a relative phase shift between right and left circular polarization components, \cite{budker2002resonant} 
causing effective birefringence for near-resonant light, which can be observed as Faraday rotation. 
Faraday effects in polarization structured light fields \cite{Stern2016} allow a high degree of freedom in controlling circular birefringence and circular dichroism of the atomic medium, enabling e.g.\ the demonstration of an all-optical isolator for radially polarized light. 

Overall, the combination of the mature research of hot and cold
atomic vapours with today's highly versatile generation of structured vector light fields offers new opportunities in light-matter interaction, shaping the spin alignment of the atomic medium and in turn the polarization structure of a transmitted optical beam.


\section{Propagation in a nonlinear medium} \label{sectionPropagation}


Atomic gasses can be understood as a nonlinear medium, where the response of the atoms to the optical field generates a complex polarizability, in turn acting as a complex refractive index for the passing optical field and causing absorption and dispersion. This can be tested with a probe beam, as discussed in the previous section, but also changes the propagation of the light field that has induced these effects in the first place. 

If the light is far enough detuned, absorption can be neglected, and the atoms can be treated as a two-level system, with a saturable self-focussing nonlinearity (exhibiting Kerr lensing). In this regime, the atoms act as a passive nonlinear medium, similar to nonlinear crystals, such as nematic liquid crystals. \cite{izdebskaya2013self,izdebskaya2018stable} Propagation of vector light through the medium can then be described by two coupled nonlinear equations, one for each polarization component, with Kerr lensing due to intensity gradients affecting the phase of each polarization component, and saturation of the atomic transition coupling the polarization amplitudes. 

For light with sufficiently high power compared to the saturation intensity, the resulting susceptibilities (especially the refractive index) can lead to higher order nonlinear optical effects, which will modulate the characteristics of the incident light such as intensity profiles and polarization distributions. In experimental situations where a balance between dispersion and lensing is achieved, spatial solitons may form. Spatial solitons can be realised in nematic liquid crystals or materials with thermal nonlocal nonlinearity as well as in atomic media, all supported by similar theoretical frameworks.

Early theoretical investigations \cite{desyatnikov2001necklace,bigelow2002stabilization,Salgueiro2004} predicted that vector vortex solitons exhibit more stable propagation for much larger distances than the corresponding scalar vortex solitons, as the long-range nonlocal nonlinear response provides a mechanism for vortex stabilization. This was experimentally confirmed in nematic crystals, \cite{izdebskaya2012observation} comparing the propagation of scalar vortex beams ($\exp(i \varphi) \hat{\sigma}_{+}$) and vector vortex beams without net angular momentum ($\exp(i \varphi) \hat{\sigma}_{+}+\exp(-i \varphi) \hat{\sigma}_{-}$). 

An alternative approach for increasing soliton stability is the use of multiple co-propagating fields, distinguished either by their color and/or by their spatial profile. Such two-color vector vortex solitons typically consist of an incoherent superposition of a vortex component, which is by itself unstable in a nonlinear medium, and a spatial soliton. The highly non-local refractive potential induced by the spatial soliton prevents the breakup of the vortex beam, leading to guided propagation with reduced defragmentation, as proposed in \cite{xu2009vector,minzoni2009stabilization,assanto2014deflection} and experimentally confirmed in nematic crystals. \cite{Izdebskaya2015} 
Very recently these studies were generalised for higher-charge vortex solitons and vector vortex solitons, and demonstrated in lead glass with strongly thermal nonlocal nonlinearity. \cite{zhang2019higher}

Similar nonlinear effects can be realised in atomic media, where the interplay between self-focussing and intensity dependent diffraction can lead to the formation of self-trapped light beams.
This was simulated and observed experimentally for vector beams propagating through a heated Rb vapour cell. \cite{Bouchard2016} The investigation showed that homogeneously polarized light beams fragment earlier than vector vortex beams and Poincar\'e beams. 
Fragmentation is known to be induced by azimuthal modulation instabilities, which due to the self-focussing effect quickly escalate.  For beams, where the polarization structure changes across the beam profile, interference effects are less drastic. 

The interaction of intense vectorial laser light with matter gives rise to a wide range of nonlinear effects, shaping the behavior of the material and the resulting light propagation. \cite{gu2018vector}  While similar effects would be expected for atomic media, so far most experiments have been performed using liquids, crystals or fibres.  

Nonlinear interaction may not only affect the spatial amplitude, and hence beam stability, upon propagation, but also the relative phase between the different spatial modes of its vector components.  It has been shown theoretically that propagation in a nonlinear (self-focussing) medium can induce a cross-phase modulation, leading to an effective polarization rotation as analysed\cite{gibson2018control,yao2019control}, and observed in the linear \cite{gu2016varying} and nonlinear regime \cite{gu2016nonlinear} respectively.  The third-order nonlinear susceptibility was studied in anisotropic Barium Fluoride crystals. \cite{wen2019anisotropic} and a large amount of literature is available on the collapse of optical vector fields.  \cite{li2012taming,chen2016vectorial,wang2018controlling}

The eigenmodes of optical fibers, including gradient-index, step-index and hollow-core photonic crystal fibers, are vector fields, \cite{ramachandran2013optical} providing the natural choice for efficient communication links \cite{kim2019robust,fang2019cylindrical} 
and multiplexing. \cite{liu2018direct,sun2019demonstration} 
Recent experiments have demonstrated quantum cryptography based on hybrid entanglement of polarization and orbital angular momentum in a (graded index) vortex fiber \cite{sit2018quantum} and over an air-core fiber.\cite{cozzolino2019air}

The nonlinearity of optical fibres leads to mode coupling and a modification of the polarization profile upon propagation, which needs to be taken into account for mode division multiplexing of classical and quantum communication. Experimental studies have demonstrated birefringence, dispersion and intermodal nonlinear interactions, such as Raman scattering and four-wave mixing, of cylindrically polarized modes in fibers. \cite{euser2011birefringence,rishoj2013experimental,goudreau2020theory} These features can be used for nonlinear quantum squeezing, permitting the creation of continuous-variable hybrid-entangled states. \cite{gabriel2011entangling}

The highly nonlinear interaction provided by gas-filled hollow-core fiber 
is particulary suitable to achieve ultra-short pulses in the few, single or even sub optical cycle regime.  Typically, a light pulse is spectrally broadened in the hollow-core fibre and then temporally compressed. Radially polarized vector beams are particularly suited to the fiber geometry.  Their propagation dynamics has been numerically studied \cite{wang2017propagation} and experimentally observed, \cite{zhao2019spectral} and compression into the few-cycle regime has been demonstrated using a krypton filled hollow-core fiber, while still maintaining its radially polarized nature. \cite{kong2019generating}

Alternatively, high power vector vortex beams can be generated by amplifying low power vector vortex beams in nonlinear gain media, e.g. single crystal fibre amplifiers, \cite{eckerle2017high,negel2017thin} 
or by energy transfer via stimulated Raman/Brillouin scattering medium. \cite{zhu2017generation}


\section{Mode conversion of vectorial light} \label{sectionMode}
\begin{figure}
\includegraphics[width=8cm]{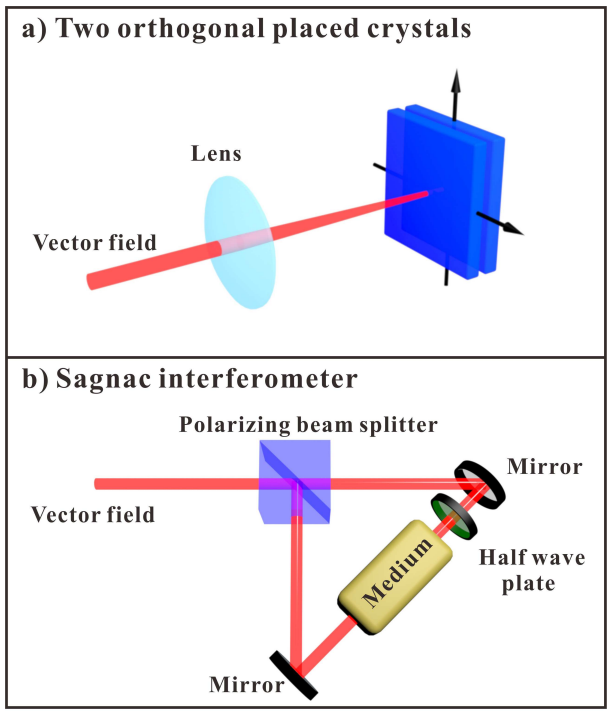}
\caption{Mode conversion of vector field, a) using orthogonal SPDC crystals subsequently addressing each polarization component, and b) in a Sagnac interferometer with counterclockwise propagating orthogonal vector components, each converted individually before recombination (the pump beam is omitted for clarity).}
\label{Modeconversion}
\end{figure}

We have already discussed how optical vector fields can be modified through linear and nonlinear effects when propagating through an atomic medium with a given susceptibility. We have also seen that the susceptibility set by the interaction with one light field can affect the propagation of another (usually weaker) field. In these situations, each of the involved light fields interacts with the atomic, or crystalline, medium through incoherent processes.  It is however also possible to drive parametric transitions, where multiple optical fields interact coherently with an atomic gas or other nonlinear medium,
and the following two sections are dedicated to this mechanism.  

Multiple-wave-mixing is a nonlinear process that coherently combines the amplitudes of some input fields to generate one or more new optical fields, mediated by the nonlinear susceptibility of the medium, and determined by the mode overlap of the participating beams. In this context, the Gouy phase associated with wavefront curvature and modenumber of the participating optical fields play important roles. As a consequence the generated signal is rigidly constrained by phase matching conditions and extremely sensitive to the polarization of the light. This makes multiple-wave-mixing an excellent candidate to implement mode conversion between spatial modes and to provide an interface for quantum networks.

An important example of nonlinear effects is harmonic generation, including second (SHG) and third harmonic generation, converting two or three photons of a pump beam into a photon at double/triple the pump frequency. A first attempt at a theoretical analysis of SHG of vectorial beams was published in 2002, \cite{freund2002second} investigating in particular the behavior of polarization singularities. For scalar vortex fields 
it can be shown that phase matching is associated with angular momentum conservation, so that the charges of the input photons' phase singularities combine, generating higher order angular momentum modes in the harmonic. For vector vortex fields the situation is more interesting:  A different phase evolution of the individual polarization components generates a polarization profile that may vary with propagation.  Moreover focused vector fields acquire an axial polarization component. This makes it more difficult to establish phase matching throughout the medium, and the efficiency of harmonic generation depends crucially on the vector structure. This was theoretically investigated for specific examples of polarization profiles, \cite{carrasco2006second,ohtsu2010calculation,grigoriev2017second} and observed \cite{kozawa2008observation} in a ZnSe crystal.

It is interesting to note that polarization always plays an implicit role in multi-wave-mixing processes, as the crystal symmetry determines the required polarization relations between input and generated beams. For type II nonlinear crystals operated in a collinear setup, the orthogonally polarized input modes can be understood as the spatial components of a vector vortex beam. The resulting multi-mode coupling has been investigated in detail for vector beams with opposite and equal OAM in the horizontal and vertical beam component, resulting in selection rules for the azimuthal and radial mode indices. \cite{pereira2017orbital} While not emphasized by the authors, this experiment demonstrated the conversion of a vector beam to a homogeneously polarized beam. Studying the effect of polarization structures in the fundamental beam on SHG was also shown to reveal varying spatial modes, while wiping out the polarization structure from the fundamental beam.\cite{zhang2018second} 

Many recent publications investigate the conversion of vector vortex beams, from fundamental infrared to visible harmonic frequencies.  The challenge here is to maintain the inhomogeneous polarization structure of the fundamental mode. This can be achieved in two complementary approaches: either by using two cascading type I crystals, \cite{liu2018nonlinear,zhang2019full,saripalli2019frequency} each addressing one polarization component of the beam at a time, or interferometrically, by separating the polarization components and performing independent SHG.
The latter has been demonstrated in a Mach-Zehnder configuration \cite{li2019nonlinear} but is more commonly performed in Sagnac interferometers. \cite{yang2019nonlinear,li2019dual,wu2020spatial}
The two abovementioned experimental setups are visually represented in   Fig.~\ref{Modeconversion}.
These experiments have shown the versatility of converting a wide range of vector beams, including polarization singularites, Poincar\'e beams and arbitrary polarization patterns. Especially the more recent work is characterized by high fidelity transfer of the vectorial mode structures, and while demonstrated so far on classical beams only, the methods should persist in the quantum regime.

Similar concepts apply also to higher-order harmonic generation, with the attractive potential of generating  extreme ultraviolet vector beams. This has been theoretically studied, \cite{hernandez2017extreme,watzel2020multipolar} and experimentally realized in a gas jet \cite{hernandez2017extreme,kong2019vectorizing} and with solid targets, \cite{kong2019spin} generating the capability of engineering XUV sources that utilise vector structures as an additional degree of freedom.

While SHG in atomic samples is forbidden due to selection rules, third harmonic generation as well as other four-wave-mixing (FWM) processes are possible. Especially FWM experiments based on quasi-resonant transitions benefit from the high efficiency of atom-light interaction compared to other nonlinear processes. Nonlinear effects in atoms are not intrinsically polarization sensitive, unless the atomic spin alignment is dictated by an external field, or a more subtle interference between Clebsch-Gordon coefficents.\cite{Walker2012} It is however possible to employ, once again, Sagnac interferometers to treat both polarization components independently in separate FWM processes, before recombining the conjugate scalar beams to a conjugate vector beam, as realised using a rubidium vapour.\cite{Hu2019} 

As a curiosity we also mention the indirect exploitation of vector beams for multiple wave mixing \cite{jiao2019multi}: passing a vector vortex beam through a polarizer results in a characteristic petal pattern with an even number of intensity lobes. When focussed into a rubidium vapour cell, they obtain different wavevectors and undergo multiple wave mixing. If individual lobes within the input pattern are blocked, they are partially regenerated through the nonlinear interaction in the atomic medium. 

Phase conjugation of vector vortex beams has also been realized in photorefractive crystals, \cite{qian2014recording,qian2014phase} demonstrating the `healing' of polarization defects, as well as in stimulated parametric downconversion. \cite{de2019real} The latter provides a convenient approach of generating a conjugate vector beam from a vectorial probe used as a seed in real time, with potential applications in aberration free imaging. Finally, we mention an experiment generating multicolor concentric vector beams, by  using cascaded four-wave mixing in a glass plate pumped by two intense vector femtosecond pulses, \cite{huang2020multicolor} combining the manipulation of temporal, spectral, spatial and polarization degrees of freedom.


\section{Electromagnetically induced transparency and memories} \label{sectionEIT}

In this section we will continue to discuss parametric light-matter interaction, but place more emphasis on the structure induced in the atoms.  In electromagnetically induced transparency (EIT) and coherent population trapping (here we will use EIT to refer to both mechanisms, which differ only in the relative intensity of the involved beams), atoms are transferred into dark states, which are determined by the interplay between the driving optical fields, and indeed their vector properties.  This can be exploited for atomic memories, where the information transferred to and stored in the atoms is retrieved at a later time.

EIT relies on the coherent interaction of light and atoms, rendering the medium transparent for a resonant probe beam when simultaneously exposed to an additional control beam. During this process, atoms are decoupled from the electromagnetic fields and populate a so-called dark state. The process can be formed by optically coupling pump and probe beams with atomic hyperfine levels or Zeeman sublevels (Hanle type resonances).  
The anomalous dispersion associated with the `transparency window' has been exploited for the generation of slow and stopped light, and related techniques have led to EIT-based quantum memories. \cite{Lvovsky2009,Hsiao2018} 
Recent research has demonstrated operation at room temperature, including EIT-based. \cite{ novikova2012electromagnetically}  
Quantum memories are a highly active research field, and here we only concentrate on a small subsection that is currently of the most relevance to applications with vector light.

Initial quantum memories dealt with information encoded in the polarization of a photon, but also spatial mode structures or images can be stored  -- with phase vortices (i.e.\ angular momentum modes) presenting a favoured basis system. A theoretical proposal to transfer and store an optical vortex in a Bose-Einstein condensate was presented in 2004. \cite{dutton2004transfer} Experimentally, the storage of optical vortices was demonstrated in atomic media at various temperatures and for classical \cite{pugatch2007topological,moretti2009collapses,firstenberg2010self,veissier2013reversible,shi2017transverse} and quantum memories. \cite{ding2013single,nicolas2014quantum,ding2015quantum,ding2016high,ding2014toward,zhou2015quantum} Image memories provide an essential capability for high-dimensional quantum memories and quantum information applications.

A different approach to high-dimensionality was demonstrated by coupling a memory for photonic polarization qubits to spatially separated output channels into multiple spatially-separate photonic channels. \cite{Chen2016memory}
Recent experiments are pushing the number of accessible spin wave modes. The simultaneous storage of up to 60 independent atomic spin-wave modes in Rb vapour \cite{Chrapkiewicz2017} and 665 spin-wave modes in a cold atomic ensemble have been demonstrated with a Raman memory, \cite{Chrapkiewicz2017} encoding modes determined by the photonic wavevector.  A similar multiplexed atomic memory system has very recently been combined with an optical cavity, coupling spin-wave excitations with different spatial profiles to the cavity photons \cite{cox2019spin} via superradiant enhancement.  
Before reviewing the storage of vector light we will highlight different parametric processes that couple optical polarization to atoms.

An intriguing application of a vector vortex beams to atoms allowed the measurement of the rotational Doppler shift, \cite{barreiro2006spectroscopic} over a decade after it was first predicted. \cite{Allen1994} The rotational Doppler shift that an atom experiences if exposed to the twisting phasefronts of an LG beam is dwarfed by linear Doppler shift along the beam propagation axis. The researchers nevertheless managed to observe the rotational Doppler broadening of the Hanle EIT signal on the D1 line of $^{87}$Rb in a room-temperature vapor cell, by using a superposition of two perfectly aligned LG fields with opposite topological charges and orthogonal circular polarizations (i.e.\ a vector beam of the form $\exp (-i \ell\varphi) \hat{\sigma}_{+} + \exp(i \ell\varphi) \hat{\sigma}_{-}$), which exactly cancels axial and radial contributions to the Doppler broadening. The ingenious idea of this experiment is to use a narrow linewidth Hanle EIT signal as a background to show the influence of the atomic rotation Doppler frequency shift induced by the topological charge $\ell$. While this work did not explicitly consider the spatial profiles or polarization distributions, it still took unique advantages of VVBs. 

\begin{figure}
\centering
\includegraphics[width=\columnwidth]{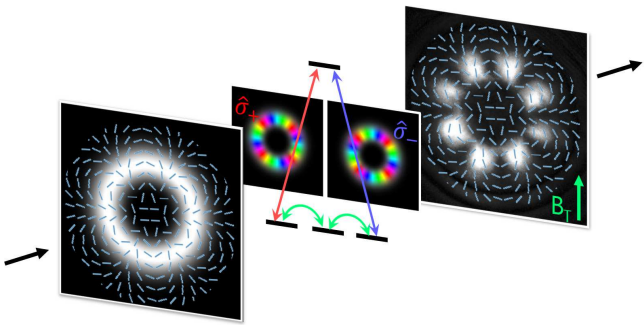}
\caption{Vector fields, in combination with an external magnetic field, result in spatially varying atomic transparency.  The orthogonal circular polarized components of a vector field (shown left) drive different transitions in an atom (center). The resulting absorption image  (right), shows transparency where the local polarization direction is aligned with the transverse magnetic field $\textbf{B}_\textrm{T}$.\label{Fig_EIT}}
\end{figure}

In 2015 we have in our group demonstrated spatially resolved EIT by exposing a cold Rb atoms to VVBs with an azimuthally varying polarization and phase structure, \cite{Radwell2015} typically with a radial polarization profile $\exp (-i \ell \varphi) \hat{\sigma}_{+} + \exp(i \ell\varphi) \hat{\sigma}_{-}$, with $\ell$ up to 200. The principle of the experiment is shown in Fig.~\ref{Fig_EIT}. The left- and right-handed circular polarization components with opposite OAM in a single laser beam provide both probe and control for the EIT transition. The atomic system is closed by a weak transverse magnetic field, making the atoms sensitive to the phase difference between the complex excitation amplitudes.  The atoms are pumped into spatially varying atomic dark states, leading to an angular variation of the opacity of the medium. The interaction allows coupling between an external magnetic field and a polarization structured optical field via the atomic spin alignment, \cite{2020AtomicCompass} where the atomic transparency is set by the angle between the local polarization direction and the external magnetic field direction \cite{selyem2019three,sharma2017phase} as shown in Fig.~\ref{Fig_EIT}. A generalisation to more complicated EIT systems has been proposed, \cite{Hamedi2018} replacing the magnetic coupling with additional optical transitions.

Related effects have also been observed in a warm vapor by using a single hybrid vector beam $\cos ( \ell \varphi) \hat{\sigma}_{+} + \sin(\ell\varphi) \hat{\sigma}_{-}$, which contains alternating segments of right and left polarized light, separated by linear polarization, and an external magnetic field in the direction of the beam propagation. \cite{Yang2019ol} Transmission of the circular light components depended strongly on the strength and direction of the magnetic field, while linearly polarized light was always absorbed. This may be interpreted as the build-up of atomic coherence through the circular polarized light. The effect prevailed when two orthogonal polarizations were provided in the form of independent homogeneous beams, and the observed transparency depended on the spatial separation between the beams. This suggests that the free flight of atoms between the two beams provides a pathway for dynamically building quantum coherence similar to adiabatic following or stimulated Raman adiabatic passage. \cite{bergmann1998coherent,bergmann2019roadmap} An alternative interpretation, following the discussion in Section \ref{sectionDichroism}, may associate the transparency with optical pumping determined by the local polarization structure and resonance conditions between the Zeeman shift atomic levels `smeared out' by the thermal motion of the atoms. 

We will now return to the realization of quantum memories for vector vortex beams.  While work on general image memories ultimately may prove to be more advantageous in accessing high dimensional spaces, the storage of vector light offers access to hybrid entanglement of polarization and angular momentum, and provides fundamental insight into the nature of light-matter interaction.
The first storage and retrieval of vector vortices at the single-photon level (with attenuated coherent light) was realized for a multiplexed ensemble of laser-cooled Cs atoms, using an intermediate configuration between EIT and off-resonant Raman schemes. \cite{Parigi2015} The polarization structure was generated with a q-plate. The authors followed a method familiar from various other approaches discussed in previous sections of treating each polarization component separately.  The vector beam was split with a calcite beam displacer into its constituent polarization components.  Each polarization components was paired with its own (homogeneous) coupling laser with the proper polarization to store the components individually in the cold atom cloud, and the retrieved signals were recombined with a second beam displacer. The overall storage and retrieval efficiency was $26\%$ for $1\,\upmu$s, and the fidelity of the atomic memory was shown to be close to one, clearly exceeding classical benchmarks for memory protocols.

More recently an EIT-based memory was realised in a warm $^{85}$Rb vapour, which is an intrinsically simpler system without additional cooling systems and beams. \cite{Ye2019} 
The vector beam, a superposition of different LG modes in the two polarization components,  was generated via a Sagnac interferometer incorporating a vortex phase plate, and split into its linear polarization components before being stored in the vapour cell. The storage and retrieval efficiency was almost $30\%$ for  $1\,\upmu$s. Projection measurements showed that both the spatial structure and phase information were preserved during storage. Quantum state tomography measurements were used to calculate the fidelities for the various VVBs. The obtained fidelities of the storage scheme in warm vapor satisfy the criterion of the quantum no-cloning theorem, offering potential for the construction of a versatile vortex-based quantum network.

Beside memories, quantum networks require various other elements, including sources of photon pairs simultaneously entangled in their polarization and spatial degrees of freedom.  Entangled vector beams combine true quantum entanglement between locally separated particles with non-classical correlations, or contextuality, between different degrees of freedom within the wavefunction of each particle. 

To the best of our knowledge, sources of entangled vector states have not yet been realised with atomic systems, but initial experiments have demonstrated just this based on SPDC in BBO crystals. \cite{DAmbrosio2016} The system works by initially generating a polarization-entangled Bell state following the usual entanglement generation, $(|h\rangle_1 |v\rangle_2 - |v\rangle_1 |h\rangle_2)/\sqrt{2} $, and passing photon 1 and photon 2 through different q-plates.  The mode profile generated by a given q-plate depends on the input polarization, and a rotation of the input polarization imposes a local phaseshift of the output polarization. A change from horizontal to vertical input relates to a change from radial to azimuthal polarization (or their higher order equivalent). The action of the q-plate therefore transfers entanglement between homogeneous polarization states to entanglement between vectorial polarization states. The authors also showed that linear local geometric transformations can generate a full Bell set, and that the states are indeed entangled and violate the CHSG inequality.

Alternatively, rather than introducing vector properties to already entangled photons, the procedure can be reversed:  the possibility of directly converting a vector beam via SPDC into hybrid entangled photon pairs has been suggested \cite{saaltink2016super} and realised. \cite{jabir2017direct} The nonlinear properties of the $\chi^{2}$ crystal can be considered for the independent polarization components of the vectorial pump beam, but specific care must be taken with phase-matching. The process of phase conjugation of vector beams was investigated theoretically for stimulated parametric downconversion. \cite{de2020quantum}

Entanglement between vector fields is not the only option when operating in a state space defined by polarization and spatial modes. Hyperentanglement has been demonstrated between the polarization of one photon and the OAM of its twin, \cite{nagali2010generation} between the polarization of one photon and a vector state of its twin, \cite{fickler2014quantum} and the connections between entanglement and contextuality have been analysed in quantum and classical settings. \cite{karimi2010spin} It is even possible to prepare the vector state of one photon by operations on its remote twin -- a vectorial analogue to ghost imaging. \cite{barreiro2010remote}  

Quantum features of hybrid entanglement have been successfully employed for several protocols, including quantum key distribution \cite{wang2019characterizing} and quantum cryptography over an outdoor free-space link. \cite{sit2017high}

\section{Light-matter interaction under strong focussing} \label{sectionFocus}

When vectorial light fields interact with matter one can distinguish two different regimes, which we call the inhomogeneous and the anisotropic regime.  In the former, the wavefunction of the atom is small compared to the spatial extent of the vector field, and individual atoms sample the local (quasi) homogeneous polarization of the light. Instead, in the anisotropic regime, an individual atom responds to the spatially varying polarization field.  This regime is particularly interesting for strongly focussed vector light, with its three-dimensional polarization structure, e.g. by positioning an atom or ion at the center of a focussed polarization vortex. The same ideas may be generalized to crystals or other matter.

Experimental work on atoms and nonlinear crystals is so far mainly situated within the inhomogeneous regime, responding to local polarization of a beam or of the spatial wavefunction of a photon, and all work described so far in this review falls into this category. Experiments with plasmons, nanowires, and individual ions and molecules instead can and have been performed in the anisotropic regime, as discussed below. These research fields are bound to benefit from the possibility to shape the 3D polarization profile. \cite{Otte2017} It is interesting to note that such polarization profiles, including Moebius strips, can be reconstructed by detecting scattering off nano-particles.\cite{bauer2015observation}  
Initial experiments have demonstrated polarization nano-tomography by using the functional nano-material itself as a sensor. \cite{otte2019polarization} 

Generally when describing the interaction between light and atoms, and specifically for all the works described so far, it is enough to describe light-matter interaction in the dipole approximation to predict the value of the observables of the system. Considering instead an atom positioned at the center of a vortex beam (where there is no intensity), the dipole approximation is not satisfied. \cite{quinteiro2017twisted} The authors devise instead a so-called Poincar{\'e} gauge, i.e.\ a gauge form of the vector potential which includes the orbital charge of the field, and evaluates explicitly the axial field component. This is further explored by proposing the use of vector light fields to generate specific electric and magnetic field components at the focus of a vector light field.\cite{Wtzel2019} At the focus of an azimuthally polarized (electric) field, without net OAM, the axial component of the electric field vanishes, whereas the axial component of the accompanying magnetic field has a finite value. The opposite applies when focussing a radial polarized light field. This allows one, in principle, to induce magnetic dipole interaction -- a regime that is usually dwarfed by the dominant electric dipole interaction. 

The strong focussing limit has been explored experimentally in interactions with single molecules. \cite{Novotny2001} 
In this work, molecules with a fixed dipole moment were used to probe the axial component of the light field, and vice versa. It was demonstrated that the orientation of single molecules could be efficiently mapped out in three dimensions by using a radially polarized beam as the excitation source. Similarly, the role of longitudinal light fields was also explored for ions, examining the interaction with focussed VVBs, \cite{quinteiro2017twisted} and the excitation of magnetic dipole transitions at optical frequencies. \cite{kasperczyk2015excitation} Another quantum system investigated under strongly focussed vector light are nanowires.  It was shown, for example, that SHG from oriented nanowires is most efficient when driven by polarization along the growth axis \cite{Bautista2015} which can be excited by illumination with strongly focused radially polarized light. 

Another route to access anisotropic interaction with vector light is to address the expanded wavefunction of Bose condensates.  Early experiments have shown that scalar vortices can be generated in quantum degenerate gasses, containing quantized angular momentum and exhibiting persistent currents.
\cite{andersen2006quantized} Vortices can also be created in spinor BEs.  The atoms of spinor BECs are in a superposition of internal quantum states, making its wave function a vector, with topological analogies to vector vortex fields.  The generation and properties of such vector vortices are the subject of ongoing investigations. \cite{wright2009sculpting,leslie2009creation,StamperKurn2013,schultz2014raman,hansen2016singular,schultz2016creating,Weiss2019}


\section{Conclusions and outlook}
The ability to generate and manipulate the vector nature of light offers new opportunities in designing light-matter interactions. This is relevant in all situations, where the symmetry of the medium differentiates between orthogonal polarization components, whether due to intrinsic dichroism or birefringence or due to assymmetries induced by external electromagnetic fields, including the vector light itself.  We have seen that, so far, many experiments deal with vector light one  vector component at a time, whether for conversion between vector modes, or for their storage. Other processes however operate more directly on a vectorial level, e.g.\ when different transitions within an atom are accessed simultaneously by the corresponding polarization components of light. One may assume that future years will see further investigations of vectorial light-matter interaction, exploring both the inhomogeneous and anisotropic nature of vector vortices as well as generic vector fields.

Current advances in the generation and detection of vector fields via miniaturization and integration for photonics devices \cite{wang2019multichannel,huang2020ultrafast,chen2020vector} in combination with new approaches based on machine learning \cite{giordani2020machine,ren2020three}
provide a new platform for technological exploitation of vectorial light-matter interaction. Photonic crystal slabs and metasurfaces offer a rich environment suitable to explore topological effects on the nanoscale and to facilitate efficient spin–orbit interaction. \cite{zhen2014topological,zito2019observation,chen2019singularities,guo2020meron,ye2020singular}

The study of complex vector fields is not restricted to photons, as is shown by recent advances in vector neutron \cite{sarenac2019generation} and electron beams. \cite{serkez2019method,morgan2020free}

We strongly suspect that vector fields have a long - and potentially twisted - future.

\section*{Acknowledgments} We are grateful for discussions with Adam Selyem on the wider research area of vectorial light matter interaction, to Alison Yao and Gian-Luca Oppo for scientific exchange on mode propagation, and to Aidan Arnold for proof-reading the manuscript. FC and SF-A acknowledge financial support from the European Training Network ColOpt, which is funded by the European Union (EU) Horizon 2020 program under the Marie Sklodowska-Curie Action, Grant Agreement No.\ 721465. JW is grateful to the China Scholarship Council (CSC) for supporting his study at the University of Glasgow through the Joint Training PhD Program (No.\ 201906280228).

\section*{Data Availability Statement} Data sharing is not applicable to this article as no new data were created or analyzed in this study.

\bibliography{refs}
\end{document}